\documentclass[twocolumn,nofootinbib,notitlepage,showpacs,prl,amsmath,amstex,amssymb,citeautoscript,longbibliography]{revtex4-2}
\pdfoutput=1
\usepackage[english]{babel}
\usepackage{letltxmacro}
\usepackage{latexsym}
\LetLtxMacro{\ORIGselectlanguage}{\selectlanguage}
\makeatletter
\DeclareRobustCommand{\selectlanguage}[1]{%
  \@ifundefined{alias@\string#1}
    {\ORIGselectlanguage{#1}}
    {\begingroup\edef\x{\endgroup
       \noexpand\ORIGselectlanguage{\@nameuse{alias@#1}}}\x}%
}
\newcommand{\definelanguagealias}[2]{%
  \@namedef{alias@#1}{#2}%
}
\makeatother
\definelanguagealias{en}{english}
\definelanguagealias{English}{english}
\usepackage{graphicx}
\usepackage{amsmath}
\usepackage{amsfonts}
\usepackage{amssymb}
\usepackage{color}
\usepackage{soul} 
\usepackage{amssymb}
\usepackage{wasysym}
\usepackage{comment}
\usepackage{dsfont}
\usepackage{xcolor}
\usepackage{float}
\usepackage{braket}
\usepackage{hyperref}
\hypersetup{
    bookmarks=false,         % show bookmarks bar?
    unicode=false,          % non-Latin characters in AcrobatÕs bookmarks
    pdftoolbar=false,        % show AcrobatÕs toolbar?
    pdfmenubar=true,        % show AcrobatÕs menu?
    pdffitwindow=false,     % window fit to page when opened
    pdfstartview={FitH},    % fits the width of the page to the window
    pdftitle={Propagation of Many-body Localization in an Anderson Insulator},    % title
    pdfauthor={Pietro Brighi},     % author
    pdfsubject={},   % subject of the document
    pdfcreator={},   % creator of the document
    pdfproducer={}, % producer of the document
    pdfkeywords={many-body localization} {MBL proximity effect} {disordered systems}, % list of keywords
    pdfnewwindow=true,      % links in new window
    colorlinks=true,       % false: boxed links; true: colored links
    linkcolor=black,          % color of internal links (change box color with linkbordercolor)
    citecolor=blue,        % color of links to bibliography
    filecolor=magenta,      % color of file links
    urlcolor=blue           % color of external links
}
\begin{document}
\title{Propagation of Many-body Localization in an Anderson Insulator}
\author{Pietro Brighi$^1$}
\author{Alexios A. Michailidis$^1$}
\author{Dmitry A. Abanin$^2$}
\author{Maksym Serbyn$^1$}
\affiliation{$^1$IST Austria, Am Campus 1, 3400 Klosterneuburg, Austria}
\affiliation{$^2$Department of Theoretical Physics, University of Geneva, 24 quai Ernest-Ansermet, 1211 Geneva, Switzerland}
\date{\today}

\begin{abstract}
Many-body localization (MBL) is an example of a dynamical phase of matter that avoids thermalization. While MBL phase is robust to weak local perturbations, the fate of an MBL system coupled to a thermalizing quantum system that represents a ``heat bath'' is an open question that is actively investigated theoretically and experimentally.  In this work we consider the stability of an Anderson insulator with a finite density of particles interacting with a single mobile impurity --- a small quantum bath. We give perturbative arguments that support the stability of localization in the strong interaction regime. Large scale tensor network simulations of dynamics are employed to corroborate the presence of the localized phase and give quantitative predictions in the thermodynamic limit. We develop a phenomenological description of the dynamics in the strong interaction regime, and demonstrate that the impurity effectively turns the Anderson insulator into an MBL phase, giving rise to non-trivial entanglement dynamics well captured by our phenomenology.
\end{abstract}

\maketitle

\textit{Introduction.---}
The thermalization of isolated quantum systems and its breakdown are fundamental questions of many-body quantum physics. While typical interacting quantum systems are expected to thermalize according to the eigenstate thermalization hypothesis (ETH)~\cite{Deutsch1991,Srednicki1994}, the presence of strong disorder can lead to non-ergodic behavior, a phenomenon known as many-body localization (MBL)~\cite{Gornyi2005a,Basko2006}. MBL and its phenomenology were thoroughly studied both theoretically~\cite{Bardarson2012,Altman2015,Nandkishore2015b,Gornyi2018,Goremykina2019,Huse2019,Abanin2019} and experimentally~\cite{Schreiber2015,Choi2016,Chiaro2019,Greiner2019a,Greiner2019b}. In one dimension, MBL is believed to be a stable phase of matter, as its stability is proven on the mathematical level of rigor for a particular model at strong disorder~\cite{Imbrie2016} under very mild assumptions. However, there are important open questions regarding the nature of MBL, such as its stability in higher dimensions~\cite{DeRoeck2017,Choi2016,Altman2019,Pal2019,Eisert2020,Mirlin2020} and the possibility of  localized  and thermalizing eigenstates coexisting in the same system at different energy or particle densities --- so-called many-body mobility edges~\cite{Luitz2015,DeRoeck2016,Guo2019,Brighi2020,Yao20,Pomata,Zakrzewski2020}.

In this work we explore the stability of MBL in the presence of a coupling to a thermalizing quantum system, dubbed as ``quantum bath". This question was addressed in various different settings at a theoretical level~\cite{Huse2015,Luitz2017,Pollmann2019,Crowley2020,Nandkishore2015a,Hyatt2017,Goihl2019,BarLev2021} and experimentally~\cite{Rubio-Abadal2019,Leonard2020}. The general expectation is that the quantum bath facilitates energy exchange and transport in the MBL system, leading to delocalization. This may be a natural outcome when the quantum bath has many more degrees of freedom compared to the MBL system or is coupled in a non-local way~\cite{Huse2015,Luitz2017,Pollmann2019,Crowley2020}. Conversely, when the bath consists of just a few degrees of freedom and is coupled locally to the MBL system it may itself localize instead of delocalizing the system, a phenomenon dubbed MBL-proximity effect~\cite{Nandkishore2015a}. Recent experiments involved a setup in which MBL and bath are represented by two species of mutually interacting quantum particles, one being subject to disorder and the other being ``clean''~\cite{Rubio-Abadal2019,Leonard2020}. Although Ref.~\cite{Rubio-Abadal2019} suggests the existence of a regime where the bath is too ``small'' and cannot delocalize the MBL particles, the experiment did not probe the fate of the bath under the back-action of the disordered particles. A related setup, although with both species of particles subject to disorder, was theoretically considered in Ref.~\cite{Krause2021}, which suggested that depending on the localization length, a bath consisting of a single boson can either localize or thermalize other particles. 

In this work, inspired by recent experiments~\cite{Rubio-Abadal2019,Leonard2020} we investigate the behavior of a non-interacting Anderson insulator with a finite density of particles interacting with a single clean (not subject to disorder potential) boson that acts as a small quantum bath. We obtain a perturbative condition for the stability of localization, which suggests that at strong disorder and strong interactions, the system remains localized. In addition, we use large scale matrix product state (MPS)~\cite{Verstraete2006} simulations to study the long-time dynamics of the system. We confirm that the Anderson insulator localizes the clean boson and propose an effective description for its dynamics. Next, we address the effect of the clean boson on the Anderson insulator, showing that by mediating an effective interaction among localized particles it triggers additional relaxation of the density imbalance~\cite{Schreiber2015}  enhancing particle-particle correlations. Finally, we attribute the non-trivial entanglement dynamics observed in this system to an effective \textit{propagation of MBL} in the Anderson insulator mediated by the interaction with the clean boson.
  
 \begin{figure*}[t]
\includegraphics[width=1.999\columnwidth]{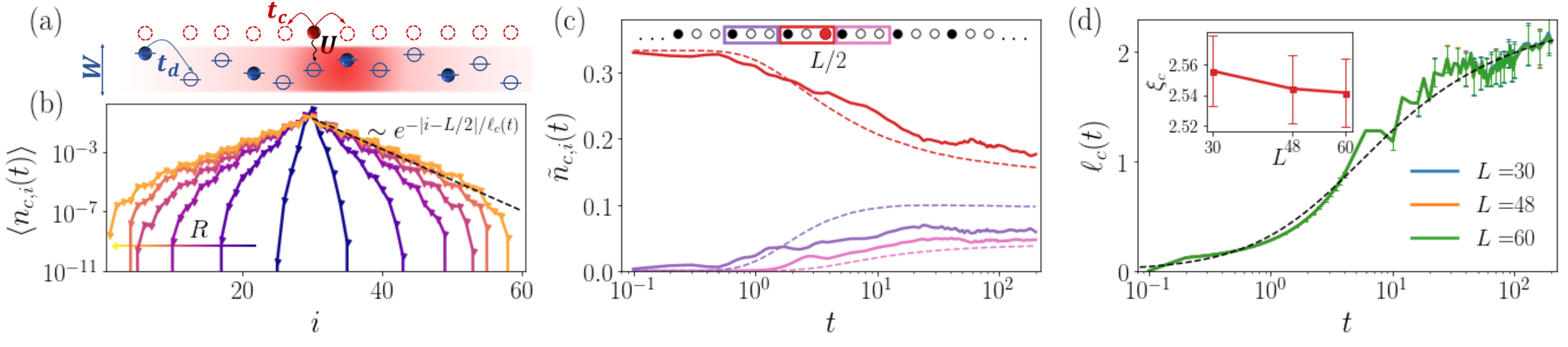}
\caption{\label{Fig:nc}
(a) Setup: a chain with finite density of $d\text{-}$bosons subject to disorder interacts with a single $c\text{-}$boson. 
(b) The density profile of $c\text{-}$boson at different times reveals an exponential decay with distance from its initial position within a region of size $R(t)$ with decay length $\ell_c(t)$, both changing with time. Color intensity corresponds to time from $t=0.4$ (darkest) to $t=200$ (brightest). 
(c) Dynamics of the coarse-grained density $\tilde n_{c,i}$ reveals saturation and agrees reasonably well with the ansatz~(\ref{Eq:nc normalized}). 
(d) The decay length of $c\text{-}$boson density profile extracted from fit to Eq.~(\ref{Eq:nc normalized}) shows a logarithmic increase for $t\geq 1$ and eventually saturates to a finite value. Dynamics of $\ell_c(t)$ agrees with ansatz~(\ref{Eq:xi c ansatz}) shown by dashed line. The inset shows the saturation value of $\ell_c(t\to\infty)$ extracted from fit to ansatz that remains constant between different system sizes. 
The data shown here are obtained averaging over $50$ disorder realizations and error bars correspond to standard deviation.
}
\end{figure*} 
  
\textit{Model.---}
We consider a chain of two species of hard-core bosons with nearest neighbor hopping: disordered ($d\text{-}$bosons) are subject to a random potential $\epsilon_i\in[-W,W]$ and interact with a single clean boson ($c\text{-}$boson) via the on-site density-density interaction with coupling $U$, see Fig.~\ref{Fig:nc}(a). The Hamiltonian
\begin{multline}
\label{Eq: H}
\hat{H} = t_d\sum_{i=1}^{L-1} \bigr(\hat{d}^\dagger_{i}\hat{d}_{i+1} +\text{H.c.}\bigr) + \sum_{i=1}^L\epsilon_i\hat{n}_{d,i} \\
+ t_c\sum_{i=1}^{L-1} \bigr(\hat{c}^\dagger_{i}\hat{c}_{i+1} +\text{H.c.}\bigr) + U\sum_{i=1}^L \hat{n}_{c,i}\hat{n}_{d,i},
\end{multline}
conserves the total number of $c$ and $d\text{-}$bosons, $[H, N_{d/c}] = 0$ where $N_{d/c} = \sum_{i}n_{d/c,i}$ and $n_{d,i} = d^\dagger_id_i$. In what follows, we assume a finite density of $d\text{-}$bosons, $\nu_d = N_d/L$. In absence of the $c\text{-}$boson the system is  in the Anderson insulating phase for any finite disorder. The presence of the $c\text{-}$boson makes the system inherently interacting, leading to the question of stability of the Anderson localization with respect to coupling to the $c\text{-}$boson.

\textit{Perturbative stability of localization.---}%
We explore the stability of localization starting from the strongly interacting limit $t_{c},t_{d} < W \ll U$. When $U\rightarrow \infty$ both species are localized since  the disordered bosons act as impenetrable walls for the clean bosons and vice versa, while the $c$-$d$ doublons form stable localized quasiparticles~\cite{Nandkishore2015a}. 
To obtain the criterion for localization at a large but finite $U$ we estimate the probability of resonant tunneling of $d$-bosons triggered by a $c\text{-}$boson. We take the basis of Anderson localized orbitals for $d$-bosons and use a mean-field (Hartree) approximation to derive the effective disorder and corresponding Anderson localized orbitals for the $c\text{-}$boson. The approximation consists of replacing $\hat{n}_{d,i}\to\langle n_{d,i}\rangle_{t\rightarrow \infty}$ 
calculated from diagonal ensemble~\cite{Polkovnikov2011}, which turns the last term of Eq.~(\ref{Eq: H}) into an effective disorder potential for $c\text{-}$bosons, $V_\text{eff} = U \sum_{i} \langle n_{d,i}\rangle_{t\rightarrow \infty} n_{c,i}$. The resulting disorder induces the localization of the $c\text{-}$boson in this approximation, with a localization length that depends on the effective disorder strength $\tilde{W}$ quantified by the variance of $U \sum_{i} \langle n_{d,i}\rangle_{t\rightarrow \infty}$. $\tilde W$ is a non-monotonic function of the disorder strength $W$ imposed on $d\text{-}$bosons, and also depends on their initial state~\cite{Brighi2021}. The localization length $\xi_c$ for $t_c\! =\!t_d\! =\! 1$ and $W\! =\! 6.5$ is much larger than lattice spacing, $a$, and the localization length of $d\text{-}$bosons, $\xi_d$, and it can be estimated~\cite{Thouless1972} as $\xi_c\propto \tilde{W}^{-2}\propto U^{-2}$.

To quantify the stability of localization, we compare the matrix element for simultaneous tunneling of $d\text{-}$ and $c\text{-}$boson triggered by the interaction $M_{cd}\approx U/(4\xi_c)$ to the effective two-particle level spacing $\delta_{cd}$~\cite{Brighi2021}. The effective level spacing is given by the number of states $\mathcal{N}$ coupled by the interaction, $\delta_{cd}  \sim  {(W+ \tilde{W})}/{\mathcal{N}}$. The criterion $M_{cd} < \delta_{cd}$ suggests that such tunneling processes are rare. Substituting the values of $
\delta_{cd}$ and $M_{cd}$ we obtain
\begin{equation}
\label{Eq:U cr strong U}
(W+\tilde{W})U>\frac{3}{2}\frac{\xi_c^0\xi_d\nu_d(1-\nu_d)}{a^2},
\end{equation}
where $\xi^0_c\!=\!U^2\xi_c$ is a rescaled $c\text{-}$boson localization length that has a finite limit when $U\to 0$.
Since the left hand side of the inequality grows as the coupling $U$ is increased, this criterion implies the stability of localization for sufficiently strong values of $U$.

\textit{Numerical simulations.---}%
To quantify the validity of the criterion~(\ref{Eq:U cr strong U}), we numerically study dynamics of large systems for long times. We set the hopping of both species $t_c=t_d=1$ and fix the disorder strength to $W=6.5$, which is close to the value that maximizes the effective disorder $\tilde{W}$~\cite{Brighi2021}. We consider a quantum quench from an initial period-3 density wave of $d\text{-}$bosons corresponding to density $\nu_d=1/3$, a typical setup used in cold atoms experiments on localization~\cite{Bloch2016,Bloch2017a,Bloch2017b}.  Given the large local Hilbert space dimension, exact diagonalization is restricted to very small systems. To access larger chains we developed a parallel version of the TEBD algorithm~\cite{Vidal2003} applied to the MPS representation of the wave function using the ITensor library~\cite{itensor}. This allowed to reach states with bond dimension up to $\chi=3000$, enabling us to explore large systems (up to $L=60$) on experimentally relevant timescales of $T=200$ hopping times.

\textit{Localization of the clean boson.---}%
We first analyze the dynamics of the $c\text{-}$boson that is prepared at the center of the chain for strong interaction $U=12$ so that Eq.~(\ref{Eq:U cr strong U}) is satisfied and we expect the $c\text{-}$boson to localize.
 Indeed, the density profiles $\langle{n_{c,i}(t)}\rangle$ shown for different times in Fig.~\ref{Fig:nc}(b) have a clear exponential decay, characterized by a time-dependent decay length $\ell_c(t)$. However, this behavior is limited to a region $R(t)$ beyond which we observe a Gaussian decay of the density~\cite{Brighi2021}. The Gaussian density profile could be interpreted as a signature of diffusive spreading within an exponentially localized profile, and indeed $R(t)$ grows approximately like $\sqrt{t}$~\cite{Brighi2021}. At the same time, the decay of the $c\text{-}$boson density away from its original position becomes exponential, signaling localization. An accurate description of the density profile for the considered range of times is given by an ansatz 
\begin{equation}
\label{Eq:nc normalized}
n_c(x,t) \approx \mathcal{N}_c(t) \exp\Bigr(-\frac{|x|}{\ell_c(t)\tanh\bigr(\frac{R(t)}{|x|}\bigr)}\Bigr),
\end{equation}
that depends on two parameters, $\ell_c(t)$ and $R(t)$ and smoothly interpolates between an exponential decay for $x\ll R(t)$, and a Gaussian decay with $\sigma^2=\ell_c(t)R(t)$ when instead $x\gg R(t)$. $\mathcal{N}_c(t)$ is a normalization factor that depends on values of $\ell_c(t)$ and $R(t)$. To reduce the effect of the period-3 density wave structure originating from the initial state of $d\text{-}$bosons, we coarse-grain the density over a unit cell of three sites $\tilde{n}_{c,i}(t) = \frac{1}{3}\sum_{j=i}^{i+3}\langle \hat{n}_{c,j}(t)\rangle$ and fit its real-space profile to Eq.~(\ref{Eq:nc normalized}) to extract the time-dependent decay length and $R(t)$. Figure~\ref{Fig:nc}(c) shows that the prediction of the ansatz (dashed lines) is adequately describing $\tilde{n}_c(i,t)$. The dynamics of the coarse-grained density also highlights the saturation of the $c\text{-}$boson dynamics to a non-thermal value, as indicated by the plateaus, thus confirming its localized nature and suggesting that both $R(t)$ and $\ell_c(t)$ eventually saturate.

The saturation of $R(t)$ is due to the finite size of the system, which limits its growth to $L/2$~\cite{Brighi2021}. On the other hand the behavior of the decay length is non-trivial. The plateaus in $\tilde{n}_c(i,t)$ and the slowdown of the growth of $\ell_c(t)$, arising around time $t\approx 10$, suggest a saturation of $\ell_c(t)$ to a finite value $\xi_c = \ell_c(t\to \infty)$ that we take as a proxy for the localization length of $c\text{-}$boson. In Figure~\ref{Fig:nc}(d) we observe a logarithmic growth of $\ell_c(t)$ that seems to saturate at later times. In order to extract the saturation value of $\ell_c$ we fit its time dependence to the following ansatz, 
\begin{equation}
\label{Eq:xi c ansatz}
\ell_c(t) = \xi_c \frac{\log\left(1+t/{T_0}\right)}{1+\log\left(1+t/T_0\right)},
\end{equation}
where $T_0$ is a characteristic saturation time. 
The numerical results are in good agreement with the fit to Eq.~(\ref{Eq:xi c ansatz}), shown as a black dashed line in Fig.~\ref{Fig:nc}(d), allowing to extract the localization length, shown in the inset of Figure~\ref{Fig:nc}(d) for $U = 12$. The localization length is constant in $L$ within error bars, suggesting that the system remains localized in the thermodynamic limit.

\begin{figure}[t]
\includegraphics[width=.95\columnwidth]{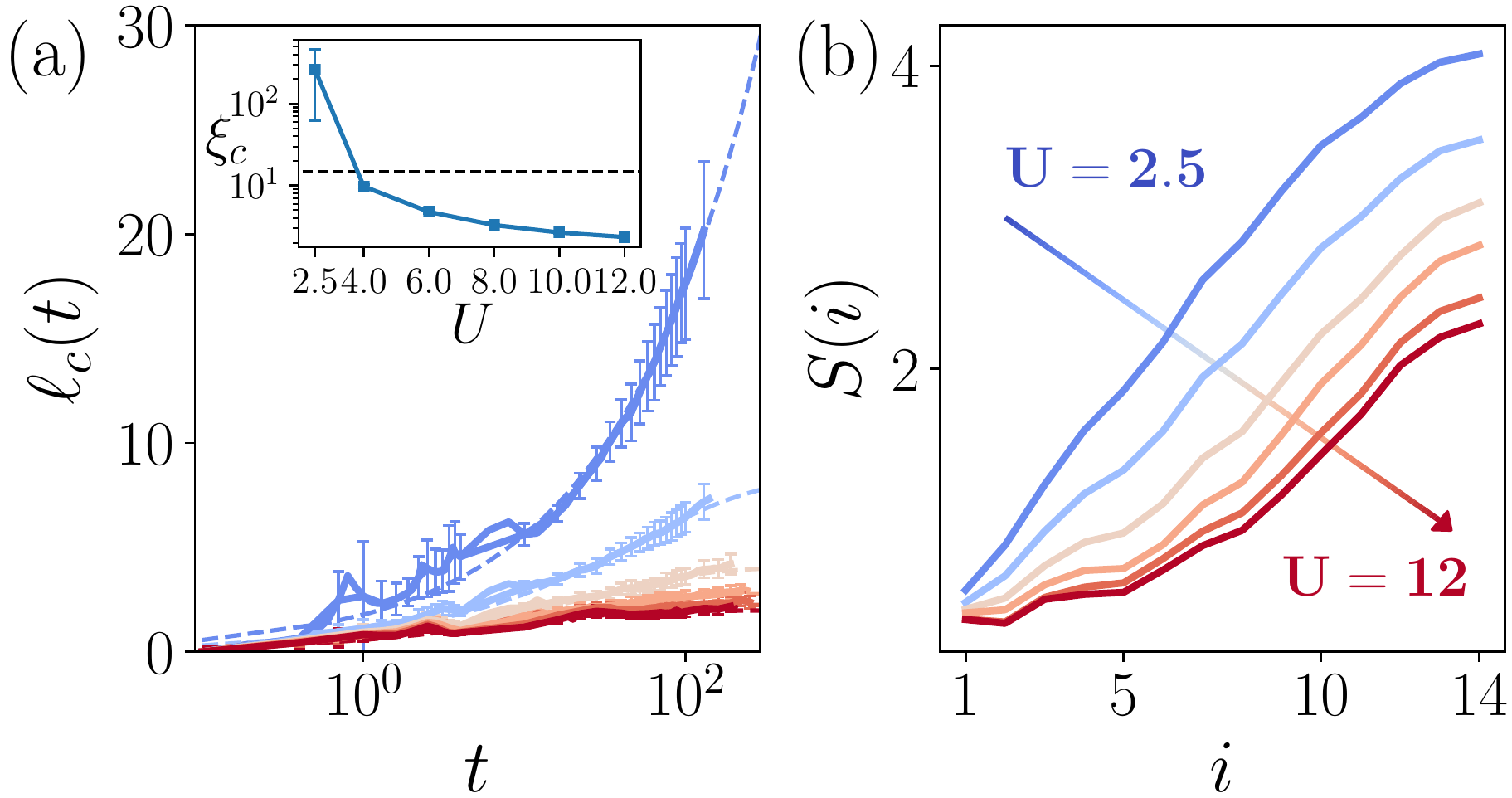}
\caption{\label{Fig:U scaling}
(a) Upon decreasing the interaction strength $U$ the growth of decay length $\ell_c(t)$ accelerates and its time dynamics becomes faster than logarithmic for $U=2.5$. Around the same value of $U$ the asymptotic value of $\ell_c(t)$ becomes larger than system size as shown in inset. Dashed lines correspond to the fit to Eq.(\ref{Eq:xi c ansatz}).
(b) The entanglement profile at time $T=130$ rapidly increases with decreasing $U$ and becomes concave for $U=2.5$ suggesting thermalization. The data are for $L=30$ averaged over $50$ disorder realizations.
}
\end{figure}

Figure~\ref{Fig:U scaling} considers the dynamics of decay length of the $c\text{-}$boson depending on the interaction strength. As expected, we observe that weaker interactions lead to much faster increase of $\ell_c(t)$, with its growth  becoming faster than logarithmic between $U=2.5$ and $U=4$. Moreover, the saturation value of decay length [inset of Fig.~\ref{Fig:U scaling}(a)] for $U=2.5$ is larger than system size, indicating that the available system size is insufficient to determine the fate of the $c\text{-}$boson at this interaction.  In addition, we consider the bipartite entanglement entropy $S(i) = -\mathop{\rm tr} \rho_i\ln \rho_i$, where $\rho_i$ is the reduced density matrix of the first $i$ sites, at fixed time $T=130$. The large values of $S(L/2)$ for weak interactions in Fig.~\ref{Fig:U scaling}(b) highlight the increasing complexity of simulations. In the remainder of the present work we focus on analyzing the regime of $U=12$, reserving the study of weaker interactions for future work.

\begin{figure}[b]
\includegraphics[width=.99\columnwidth]{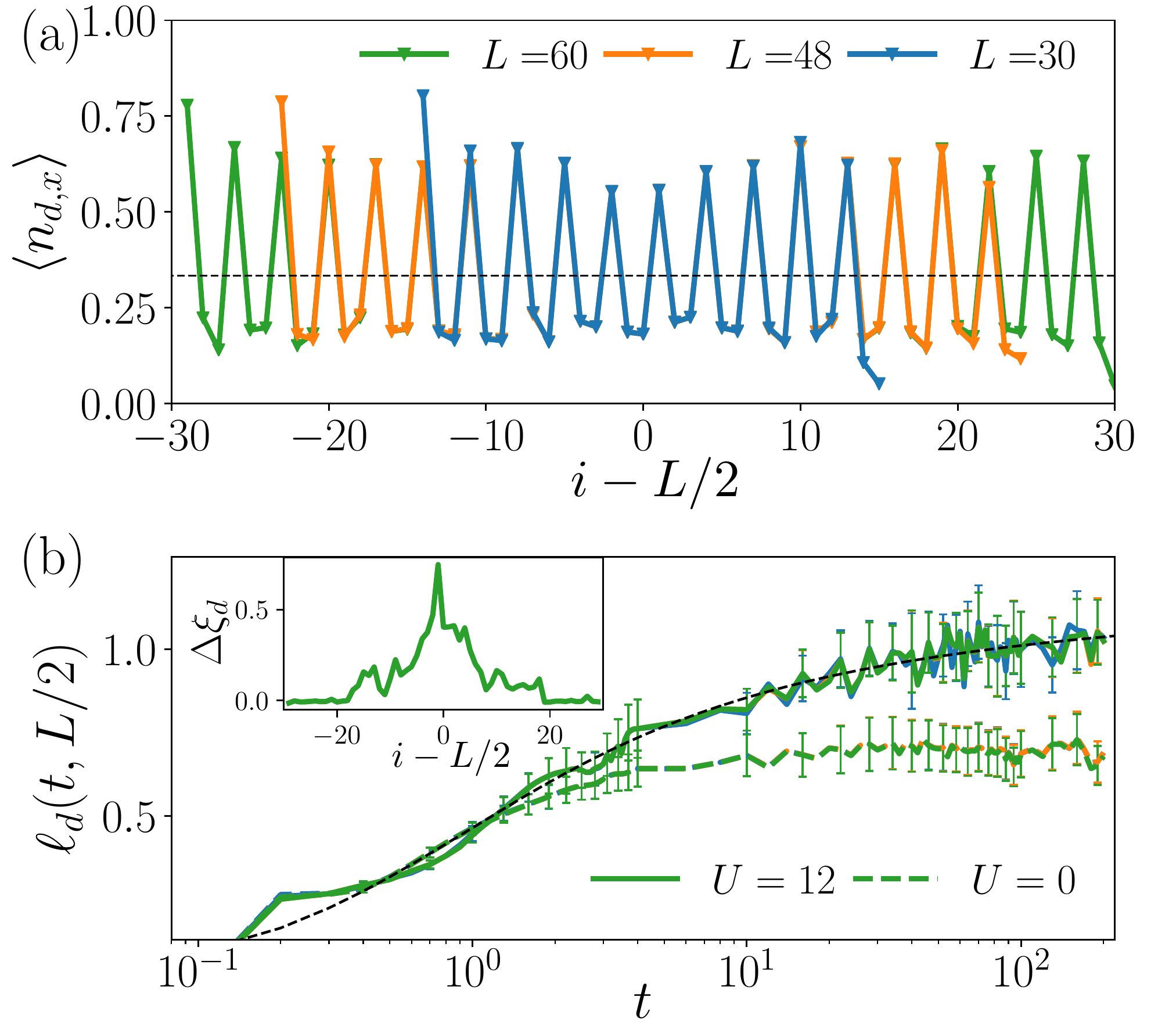}
\caption{\label{Fig:nd}
(a) The density profiles at late times $T=200$ show memory of initial state of $d\text{-}$bosons. The $c\text{-}$boson leads to an additional small relaxation of density only near the center of the chain. 
(b) The decay length of $d\text{-}$bosons also reveals logarithmic in time dynamics and saturates even in the middle of the chain. Inset: the difference between saturation values of $\ell_d(t\to\infty,i)$ for $U=12$ and $U=0$ extracted from Eq.~(\ref{Eq:xi c ansatz}) shows enhancement of the decay length in the central region compared to Anderson insulator.
}
\end{figure}

\textit{Localization of $d\text{-}$bosons.---}%
We proceed with the characterization of the effect of the $c\text{-}$boson on the localized $d\text{-}$bosons for large values of $U$. In order to analyze the dynamics of the $d\text{-}$bosons, we study the memory of their initial state in Fig.~\ref{Fig:nd}(a), where we show the density profiles at late time $T=200$ for system sizes $L=30,48,60$. The initial state is a period-$3$ density wave and the $\langle n_{d,i}\rangle$ curves immediately suggest that the time evolution did not wash out the initial density wave structure. This result provides a clear sign of absence of thermalization, at least on the timescales considered. We notice however an enhanced relaxation close to the center of the chain. This is readily explained by the interaction with the $c\text{-}$boson ($i=L/2$) that remains localized in the central region of the chain.

The presence of the localized $c\text{-}$boson induces effective interaction between $d\text{-}$bosons. The effect of these interactions can be probed via the connected correlation functions $\langle n_{d,i}(t)n_{d,j}(t)\rangle_c= \langle n_{d,i}(t)n_{d,j}(t)\rangle - \langle n_{d,i}(t)\rangle\langle n_{d,j}(t)\rangle$. The exponential decay of the absolute value of the disorder-averaged connected correlation function with distance $x$, $|{{\langle n_{d,i-x/2}(t)n_{d,i+x/2}(t)\rangle_c }}|{\sim} e^{-\frac{x}{\ell_d(t,i)}}$ allows to define a time and position dependent decay length of $d\text{-}$bosons, $\ell_d(t,i)$. We use the asymptotic value of $\ell_d(t\to\infty,i)$ at large times as a proxy for the $d\text{-}$bosons localization length~\cite{Brighi2021}. Given the inhomogeneity of the system, we expect $\ell_d(t,i)$ to be different, depending on the distance from the middle of the chain. This expectation is confirmed by Fig.~\ref{Fig:nd}(b) that reveals that the decay length is largest at $i=L/2$ and decreases away from the initial position of the $c\text{-}$boson. Although the value of $\ell_d$ is larger than the Anderson value (dashed lines), it remains close to one lattice spacing even at late times and shows clear signs of saturation, as expected. In the inset of Fig.~\ref{Fig:nd}(b), fitting $\ell_d(t,i)$ for fixed $i$ with Eq.~(\ref{Eq:xi c ansatz}), we obtain a profile of the saturation value from which we can observe that the effect of the interaction is limited to the center of the chain. 
\begin{figure}[t]
\includegraphics[width=.95\columnwidth]{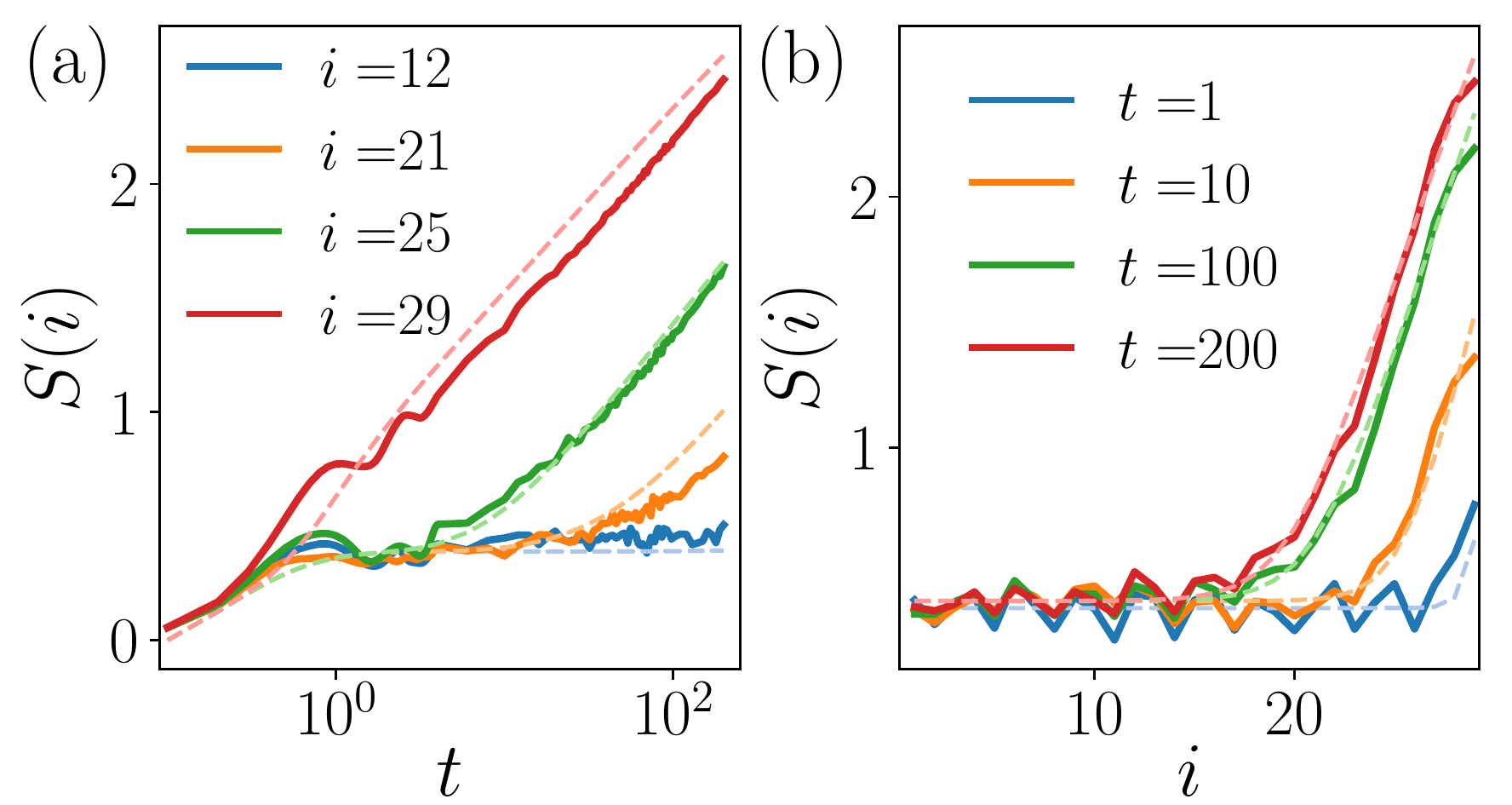}
\caption{\label{Fig:S}
(a) The logarithmic entanglement dynamics with the delayed start of the growth away from the center is naturally explained by the interactions triggered by $c\text{-}$boson. (b)~The entanglement profile at fixed times has linear growth with distance to the position of $c\text{-}$boson, $|i-L/2|$. Dashed lines in both panels are prediction from Eq.~(\ref{Eq:S ansatz}). System size is $L=60$ averaged over $50$ disorder realizations.
}
\end{figure}

\textit{Entanglement dynamics.---}%
Finally, we study the dynamics of entanglement entropy that is known to be logarithmic in the MBL phase~\cite{Serbyn2013a,Serbyn2013,Huse2014} and trivial in the Anderson insulator. In our system, despite the absence of interactions between $d\text{-}$bosons, entanglement entropy for different cuts shows logarithmic growth in Fig.~\ref{Fig:S}(a). However, the onset of this increase strongly depends on the location within the chain. We attribute such entanglement dynamics to the presence of the $c\text{-}$boson. While a single $c\text{-}$boson is incapable of producing more than $\ln 2$ entanglement, the non-zero $c\text{-}$boson density induces effective interactions between $d\text{-}$bosons thus turning the former Anderson insulator into a genuine MBL phase. 

Although the $c\text{-}$boson is eventually localized, its density is spreading throughout the chain within an exponential envelope (see Fig.~\ref{Fig:nc}(b)) effectively triggering \textit{propagation of MBL}.  Phenomenologically we describe the entanglement profile as 
\begin{equation}
\label{Eq:S ansatz}
S(i,t) = S_\text{AI}+\xi_S\log\bigr(1+U_\text{eff}(i-L/2,t)t),
\end{equation}
which captures the logarithmic growth of entanglement in the MBL phase~\cite{Serbyn2013a,Serbyn2013,Huse2014} with  an effective interaction strength $U_\text{eff}(x,t) = U n_c(x,t)$ set by the $c\text{-}$boson density from Eq.~(\ref{Eq:nc normalized}). Eq.~(\ref{Eq:S ansatz}) predicts that far from the center (large $|i-L/2|$) the logarithmic growth starts only when the \textit{many-body localization front}  $x_\text{MBL}(t)$ reaches the position of the cut. The value of $x_\text{MBL}(t)$ obtained from the condition $tU_\textit{eff}(x,t)\approx 1$ reads 
$x_\text{MBL} \approx \ell_c(t)\log\bigr(\mathcal{N}_c(t)Ut\bigr)$,
where we used that at late times $R(t)$ saturates ($\tanh(R(t)/|x|)\approx1$). The MBL front, then, at first grows as the decay length $\ell_c(t)$ and continues to grow logarithmically, even after the saturation of $\ell_c(t)$, eventually reaching the boundaries of the system. Comparison of the prediction of Eq.~(\ref{Eq:S ansatz}) with data in Fig.~\ref{Fig:S}(a) shows good agreement. Furthermore, from Eq.(\ref{Eq:S ansatz}) the late time entanglement profile can be predicted to be linear in $|x|$, with a slope given by $-\xi_S/\ell_c(t)$. This is supported by Fig.~\ref{Fig:S}(b), where the entanglement entropy decreases linearly with the distance from the center of the chain. This allows us to interpret the curvature of entanglement profile at weak $U$ in Fig.~\ref{Fig:U scaling}(b) as a breakdown of this picture, potentially indicating delocalization. 

\textit{Discussion.---}We investigated the fate of an Anderson localized system coupled to a small local bath, represented by  a single disorder-free particle. We obtained an analytic criterion for the stability of localization for strong interactions and disorder values. The stability of localization is confirmed by unbiased MPS numerical simulations  in the fully interacting system on experimentally relevant timescales. In addition, our MPS simulations reveal potential delocalization at weak interactions. At strong interactions, when the system is localized, we propose a phenomenological picture of \textit{propagation of MBL} in the Anderson insulator, triggered by the localized $c\text{-}$boson. In particular we explain the retarded growth of entanglement far from the initial position of the $c\text{-}$boson and the enhanced relaxation only in the region close to it. Our predictions can be readily tested in existing experimental setups, and could be potentially extended to the case of finite $c\text{-}$bosons density. 

{\it Acknowledgments.---}We acknowledge useful discussions with M.~Ljubotina. P. B.,  A. M., and M. S. were supported by the European Research Council (ERC) under the European Union's Horizon 2020 research and innovation program (Grant Agreement No.~850899). D. A. was supported by 
the Swiss National Science Foundation and by the European Research Council (ERC) under the European Union's Horizon 2020 research and innovation program (Grant Agreement No.~864597). The development of parallel TEBD code was was supported by S.~Elefante from the Scientific Computing (SciComp) that is part of Scientific Service Units (SSU) of IST Austria. Some of the computations were performed on the Baobab cluster of the University
of Geneva.

\end{document}